\documentclass[preprint,aps,showpacs,prd,tightenlines]{revtex4}
\usepackage{graphicx}
\newcommand{\be}{\begin{eqnarray}}
\newcommand{\ee}{\end{eqnarray}}
\newcommand{\nn}{~\nonumber \\}
\newcommand{\p}{{\cal P}\exp}

\newcommand{\ssh}{\gamma\cdot}

\newcommand{\bmp}{\begin{minipage}{16cm}}
\newcommand{\emp}{\end{minipage}\vskip 7mm} 

\begin{document}


\title{Fermion production in time-dependent fields}
\author{Dennis D. Dietrich}
\affiliation{
Institut f\"ur Theoretische Physik,
J.W. Goethe-Universit\"at,
60054 Frankfurt am Main,
Germany\\
and\\
Institut de Recherches Subatomiques,
23 rue du Loess
BP 28,
67037 Strasbourg Cedex,
France
}

\date{October 27, 2003}


\begin{abstract}

The exact fermion propagator in a classical time-dependent gauge field is
derived by solving the equation of motion for the Dirac Green's functions.
From the retarded propagator obtained in this way the momentum spectrum for
the produced fermion pairs is calculated. Different approximations and the
exact solution for the propagator and the momentum spectrum are presented.

\end{abstract}



\maketitle


\section{Introduction}

Particle production in classical bosonic fields has been a topic of continuing
interest in quantum electrodynamics (QED) and quantum chromodynamics (QCD). 
It is relevant for the physics of the early universe \cite{cornwall}, of
intense laser fields \cite{laser} as well as 
of ultra-relativistic heavy-ion collisions and the quark-gluon plasma 
\cite{qgp} (QGP). A lot of effort is made to study the QGP's production and 
equilibration \cite{kinetic} in nuclear collision experiments at the 
Relativistic Heavy-Ion Collider (RHIC) at the Brookhaven National
Laboratory (BNL) and the Large Hadron Collider (LHC) under construction at
CERN. The existence of such a state of matter is predicted by lattice QCD 
calculations at high temperatures \cite{lattice}.

At ultra-relativistic energies, the two nuclei are highly Lorentz
contracted. When they pass through each other, a chromoelectric field is
formed due to the exchange of soft gluons \cite{soft}. This is a natural 
extension of the color flux-tube model or the string model which are widely 
applied to high-energy $pp$, $e^+e^-$, and $pA$ collisions \cite{string}. 
Many other recent publications, e.g. \cite{initial,epsilon,alpha,authors}
are based on the hypothesis that the initial state in heavy-ion 
collisions is dominated by
gluons which on account of the large occupation number can be treated as a
classical background field.

The larger the occupation number $\left<c_k^\dagger c_k\right>$ of the bosonic
sector of a physical system, the better it can be described by a
classical field. Here $c_k$ and $c_k^\dagger$ are the bosonic field
annihilation and creation operators for particles of momentum $k$. Especially 
if the occupation number is much larger than one, the commutator of the 
creation and annihilation operator and hence quantum effects can be neglected:
$[c_k^\dagger,c_k]=1\ll \left<c_k^\dagger c_k\right>$. 
The field operators can be approximated by complex numbers, i.e. they are 
treated classically.

For gluons in a heavy-ion collision at RHIC with $\sqrt{s}=130$GeV the 
initial occupation number for gluons of average transverse momentum 
$|\vec k_T|\approx 1$GeV in the center of the collisions is roughly equal to 
1.5 \cite{mueller}. Although this number is not much larger than unity, the
classical field as the expectation value of the gauge field still
constitutes the main contribution as compared to the fluctuations of the
gauge field. The occupation number for lower transverse momenta is yet
higher and the classical concept is an even better approximation. For
larger
transverse momenta the occupation number is smaller and thus the quantum
fluctuations are more important there. As the above value for the occupation
number at the average transverse momentum is larger than one, keeping only 
classical bosons is justified as a first approximation. One can investigate 
quantum fluctuations in a subsequent step.

The high-occupation number bosonic fields are of the order 
$A\sim g^{-1}$. Thus processes with multiple couplings to the classical field
are not parametrically suppressed by powers of the coupling constant $g$.
Without an additional scale, they have to be taken into account to all
orders. Under the prerequisite of weak coupling, the most important quantum
processes involve only terms of the classical action which are of second
order in the quantum fields. These are the fermion and the antifermion
fields as well as the field of the bosonic quantum fluctuations. The
coefficient of the second order terms for a given field constitutes
the inverse of the corresponding two-point Green's function. Inversion for 
selected boundary conditions yields the particle's propagator. The
propagators contain all the information on two-(quantum-)particle reactions in 
the presence of classical fields to all orders in the coupling constant $g$. 
These reactions are scattering off the classical field or particle production 
by vacuum polarisation. 

In the following let us consider particle production. In quantum
electrodynamics this means production of electron-positron pairs. 
Analogously, in quantum chromodynamics quark-antiquark pairs can be produced. 
However, due to the non-linearity of the field
tensor pairs of gluonic quantum fluctuations are produced, too. In fields of 
the magnitude $A\sim g^{-1}$ the production of both kinds of pairs is
equally parametrically favoured. This paper only deals with the production of 
fermions and antifermions.
It is possible that in a given situation the bosonic sector is covered
by the concept of a classical field sufficiently well. Corrections to the
high-momentum sector could be perturbatively accessible \cite{ddd}.

No concept of a classical field exists for the fermionic sector. There the
occupation number has always to be less than unity due to the Pauli
principle. A purely perturbative treatment could only describe the
high-momentum sector. The soft part would not be treated consistently.

An alternative to the perturbative approach for particle production is
Schwinger's constant-field method \cite{schwinger} which is an exact
one-loop non-perturbative approach. This method can also be understood as
semiclassical tunneling across the mass gap \cite{cnn}. However, this scheme
is based on the assumption of a slowly varying classical field. If the field
changes too rapidly in space or time, the production of fermions is again
not described properly. A different concept is needed that is independent of
energy or time scales, respectively. Such a concept is especially important,
if the time scale for a process is to be determined by a self-consistent
calculation. If an approach is applied for such an investigation, which
relies already on an assumption about the time scale, the result is likely
to be misleading. For example, if the decay time scale for a classical field
is to be calculated based on particle production, perturbative concepts are
likely to lead to times which are too short, while the Schwinger method
tends to predict a development of the system that is too slow.

For a concise treatment other methods are necessary. Exact results are
desirable but hard to obtain. As mentioned before, neglecting bosonic
quantum fluctuations the behaviour of the fermions is governed by their
two-point functions in the classic background field. It can be obtained by
solving the equation of motion for the Dirac
Green's function $G(x,y)$ exactly:

\be
[i\ssh\partial(x)+\ssh A(x)-m]G(x,y)=\delta^{(4)}(x-y).
\label{dirac}
\ee

There are other ways to derive the full propagator, for instance by 
resumming all terms of the perturbative series or by adding up a set
of (at all times complete) wave-function solutions of the Dirac equation.

In arbitrary fields a few general approximations are known.  Neglecting the
field in the equation of motion leads to the free Green's function
$G^0(x-y)$. The standard perturbative series is a sum of terms containing powers
of the background field $A$ between free Green's functions. The asymptotic
behaviour of the free Green's function determines that of the approximated
full Green's function. Another approach which applies in an arbitrary field
is the static approximation $G^S(x,y)$. It is obtained by neglecting the
spatial part of the covariant derivative in the differential equation. The
remaining ordinary differential equation can be solved by direct
integration. Yet another approach can be found in \cite{gromes}.

The following investigation concerns the case where the 
classical field depends arbitrarily on one rectilinear coordinate
$A=A(n\cdot x)$. The equation of motion for the propagators shall be solved
directly. If such solutions are investigated, care has to be taken: the 
result could be any Green's function which is not necessarily a propagator. If 
a propagator has been obtained, the imposed boundary conditions determine 
whether the result is the retarded or the Feynman propagator or one of their 
related singular functions.

A solution in a field depending on one rectilinear coordinate can also be
seen as an approximation for the case where the strongest dependence is on 
this rectilinear coordinate and the dependence on all the others is much 
weaker. 

It will be explored whether an approximation scheme can be found that is 
independent of assumptions on time and/or energy scales over a large range of 
parameters. The investigated approximations are the Born, the weak-field,
the strong-field, and the Abelian approximation. The weak-field approach is an
expansion in powers of the gauge field $A$ based on the free propagator and
valid for $A\ll\omega$. $\omega$ stands for the on-shell energy of the
described particles. The strong-field approximation is justified for $\omega\ll
A$ and consists of an expansion in powers of the on-shell energy. For the
Abelian approach the commutators of the elements of the Clifford and the
charge algebra are neglected.

Chapter II includes the exact solution of the equation of motion for the
Dirac Green's function and gives several approximations to the full
solution. Section III contains the application of the previous findings to
the problem of particle production and the comparison of the different
schemes to the exact result. In the last chapter the contents of the paper
are summarised.

Throughout the paper the metric is: $g^{\mu\nu}={\rm
diag}(1,-1,-1,-1)$, angular momenta are measured in units of $\hbar$, and
velocities in fractions of the speed of light $c$. From hereon, the coupling
constant is included in the classical field: $gA^\mu_{old}=A^\mu_{new}$.


\section{Determination of the propagator}

Let us consider homogeneous solutions $G_H(x,y)$ of Dirac's equation
(\ref{dirac}).  In the special class of fields which only depend on one
rectilinear coordinate $n\cdot x$ this equation can be Fourier transformed
(three-dimensionally) into an ordinary differential equation:

\be
\left[i(\ssh n)\frac{d}{d(n\cdot x)} +\ssh\kappa+\ssh A(n\cdot x)-m\right]
G_H(n\cdot x,n\cdot y,\kappa)=0,
\label{fourier}
\ee

with the conserved three-dimensional momentum coordinate 
\mbox{$\kappa=k-n\partial(k\cdot x)/\partial(n\cdot x)$} orthogonal to $n$ 
and where $k$ stands for the four-momentum.
As a further {\it ansatz} the matrix function 
$G_H(n\cdot x,n\cdot y,\kappa)$ is to be
a functional of another matrix function $g_H(n\cdot x,n\cdot y,\kappa)$ with
a special property for the derivative:

\be
\frac{d}{d(n\cdot x)}G_H[g_H(n\cdot x,n\cdot y,\kappa)]
=
\left[\frac{d}{d(n\cdot x)}g_H(n\cdot x,n\cdot y,\kappa)\right]
G_H[g_H(n\cdot x,n\cdot y,\kappa)],
\label{functional}
\ee

which looks like the derivative of an exponential function,
but is not quite due to
the matrix structure. Provided a function $g_H(n\cdot x,n\cdot y,\kappa)$
exists which satisfies equation (\ref{functional}), the form of the functional 
$G_H[g_H(n\cdot x,n\cdot y,\kappa)]$ can be determined. Exploiting the above 
property leads to a factorisation in equation (\ref{fourier}):

\be
\left[i(\ssh n)\frac{d}{d(n\cdot x)}g_H(n\cdot x,n\cdot
y,\kappa) +\ssh\kappa+\ssh A(n\cdot x)-m\right]
\times
G_H[g_H(n\cdot x,n\cdot y,\kappa)]
=
0.
\nn
\ee

Given the existence of a solution $G_H$ other than the trivial solution, its
matrix structure can be inverted. Multiplication with the inverse of the
solution from the right then yields a differential equation for the function
$g_H$:

\be
i(\ssh n)\frac{d}{d(n\cdot x)}g_H(n\cdot x,n\cdot
y,\kappa)+\ssh\kappa+\ssh A(n\cdot x)-m
=
0.
\ee

This ordinary differential equation can be solved by direct integration where 
the initial condition $g_H(n\cdot x=n\cdot y,n\cdot y,\kappa)=0$ is chosen:

\be
g_H(n\cdot x,n\cdot y,\kappa)
=
i\frac{\ssh n}{n^2}\int_{n\cdot y}^{n\cdot x}d(n\cdot\xi)
[\ssh\kappa+\ssh A(n\cdot\xi)-m].
\label{argument}
\ee

Here it is necessary to require $n^2\neq0$. Otherwise, the
matrix $\ssh n$ does not posses an inverse because of $\det\{\ssh
n\}=(n^2)^2$. For cases with $n^2=0$ a different treatment is necessary.

In general, the argument function $g_H$ does not commute with itself at
different points $n\cdot x$. This is not (only) due to non-Abelian charges
which might be included in the vector field $A$ but to the non-commutative
nature of the elements of the Clifford algebra. Thus the solution of
equation (\ref{functional}) is not an exponential function 
but a path-ordered exponential:

\be
G_H[g_H(n\cdot x, n\cdot y,\kappa)]
=
\p\{g_H(n\cdot x,n\cdot y,\kappa)\}.
\label{poe}
\ee

A sufficient but not necessary condition for its existence is that the norm
of the integrand in equation (\ref{argument}) is bounded.
It has to be noted that the invariance of an integral under simultaneous
exchange of the integration boundaries and inversion of the sign cannot be
used in equation (\ref{argument}) because the path ordering would be
reversed.

If a more general initial condition had been chosen in equation
(\ref{argument}) the additional addend $g_H(n\cdot y,n\cdot y,\kappa)$ would
have lacked an ordering parameter necessary for the path-ordering. Hence, it
could only be treated by always putting it to the rhs of the remaining
path-ordered exponential. This would have led to an extra factor
$\times\exp\{g_H(n\cdot y,n\cdot y,\vec k)\}$. As here a homogeneous
differential equation is investigated this factor does not lead to
independent solutions.

Hereafter one has only to distinguish between the cases $n^2>0$ and
$n^2<0$, because every field $A=A(n\cdot x)$ can be transformed into a
field $A=A(n'\cdot x)$ with ${\rm sgn}(n^2)={\rm sgn}({n^\prime}^2)$ by
a Lorentz transformation. Overall factors in front of the normal
vectors can be absorbed in a redefinition of the vector potential $A$. So,
for the sake of simplicity it suffices to investigate one special case per
class of fields. This is going to be done for the cases of
$n^\mu=(1,0,0,0)$ and $n^\mu=(0,0,0,1)$. In situations where $n^2=0$, 
rotations in three-space can turn any normal vector $n$ into
$n^\mu=(1,0,0,-1)/\sqrt{2}$.


\subsection{Time-like coordinates}

For a purely time-dependent field, the solution for $g_H(x_0,y_0,\vec k)$ in 
equation (\ref{argument}) is given by:

\be
g_H(x_0,y_0,\vec k)
=
i\gamma^0\int_{y_0}^{x_0}d\xi_0
[\gamma^jk_j+\ssh A(\xi_0)-m],
\label{argumentt}
\ee

with $j\in\{1,2,3\}$. Constructing the matrix function 
$G_H(x_0,y_0,\vec k)$  by putting equation (\ref{argumentt}) 
into equation (\ref{poe}) leads to:

\be
G_H(x_0,y_0,\vec k)
=
\p\left\{
i\gamma^0\int_{y_0}^{x_0}d\xi_0[\gamma^jk_j+\ssh A(\xi_0)-m]
\right\}.
\label{fullt}
\ee

In the following, various approximations are studied in order to learn more
about the above solution.


\subsubsection{Weak-field approximation}

It is useful to investigate the case of a vanishing gauge field $A=0$. One
then sees that the argument $g_H^0(x_0-y_0,\vec k)$ now commutes with itself
at different space-time points. The path-ordered exponential can now be
replaced by an exponential function.  The exponential function of matrices
can be recast into exponential functions of scalar arguments multiplied with
matrices:

\be
G_H^0(x_0-y_0,\vec k)
&=&
\gamma^0\frac{\gamma^0\omega+\gamma^jk_j-m}{2\omega}
e^{+i\omega(x_0-y_0)}
+
\gamma^0\frac{\gamma^0\omega-\gamma^jk_j+m}{2\omega}
e^{-i\omega(x_0-y_0)},
\label{freet}
\ee

with $\omega=\sqrt{|\vec k|^2+m^2}$. 

Standard perturbation theory for small gauge fields $A\ll\omega$ which can
be interpreted as an ultraviolet approximation is obtained by expanding the
exact solution in powers of $A$. Prior to this, it has to be rewritten
in order to include all powers of the momenta and the mass with every factor
of the field. The path-ordered exponential can be expressed as:

\be
G_H(x_0,y_0,\vec k)
=
\lim_{N\rightarrow\infty}
{\cal P}
\prod_{n=0}^{N-1}
\left(
1+i\gamma^0\Delta\xi^{(n)}_0[\gamma^j k_j+\ssh A(\xi^{(n)}_0)-m]
\right),
\label{poelin}
\ee 

The interval $[x_0,y_0]$ is decomposed into $N$ disjoint pieces with the
lengths $\Delta\xi^{(n)}_0$ which need not be equal and each with an inner 
point $\xi^{(n)}_0$. These are arranged according to
$x_0=\xi_0<\xi_1<...<\xi_N<\xi_{N+1}=y_0$ for $x_0<y_0$ or 
$x_0=\xi_0>\xi_1>...>\xi_N>\xi_{N+1}=y_0$ for $x_0>y_0$. ${\cal P}$
indicates that the factors are ordered with respect to the index $\nu$ where
the term with the lowest index is put furthest to the left. 
The expression can now be sorted with respect to powers
of the field $A$:

\be
G_H(x_0,y_0,\vec k)
&=&
\lim_{N\rightarrow\infty}
\sum_{l=0}^N
~\sum_{n_1=0}^{N-1}\sum_{n_2=n_1+1}^{N-1}...\sum_{n_l=n_{l-1}+1}^{N-1}
\nn
&~&
\prod_{L=0}^{n_1-1}
\left(1+i\gamma^0\Delta\xi^{(L)}_0[\gamma^j k_j-m]\right)
\times
\left[i\gamma^0\ssh A(\xi^{(n_1)}_0)\Delta\xi^{(n_1)}_0\right]
\times
\nn
&\times&
\prod_{L=n_1+1}^{n_2-1}
\left(1+i\gamma^0\Delta\xi^{(L)}_0[\gamma^j k_j-m]\right)
\times
\left[i\gamma^0\ssh A(\xi^{(n_2)}_0)\Delta\xi^{(n_2)}_0\right]
\times
\nn
&\times&
...
\times
\nn
&\times&
\prod_{L=n_{l-1}+1}^{n_l-1}
\left(1+i\gamma^0\Delta\xi^{(L)}_0[\gamma^j k_j-m]\right)
\times
\left[i\gamma^0\ssh A(\xi^{(n_{l-1})}_0)\Delta\xi^{(n_{l-1})}_0\right]
\times
\nn
&\times&
\prod_{L=n_l+1}^{N-1}
\left(1+i\gamma^0\Delta\xi^{(L)}_0[\gamma^j k_j-m]\right).
\label{resummation}
\ee

For $l=0$ there are no further sums over $n_i$. Sums and products are
not taken into account if the starting index is greater than the ending
index.
In the limit $N\rightarrow\infty$ the outer sum over the powers $l$ of the
gauge field $A$ becomes an infinite sum, the intermediate sums turn into 
integrals over simplices, and the products give path-ordered exponentials. In 
fact, their arguments commute at every point, thus the path-ordering can be
dropped here:

\be
G_H(x_0,y_0,\vec k)
&=&
\sum_{l=0}^\infty
\int_{y_0}^{x_0}d\xi_1\int^{\xi_1}_{y_0}d\xi_2...\int^{\xi_{l-1}}_{y_0}d\xi_l
\nn
&~&
\exp\left\{i\gamma^0[\gamma^j k_j-m](x_0-\xi_1)\right\}
\times
\left[i\gamma^0\ssh A(\xi_1)\right]
\times
\nn
&\times&
\exp\left\{i\gamma^0[\gamma^j k_j-m](\xi_1-\xi_2)\right\}
\times
\left[i\gamma^0\ssh A(\xi_2)\right]
\times
\nn
&\times&
...
\times
\nn
&\times&
\exp\left\{i\gamma^0[\gamma^j k_j-m](\xi_{l-1}-\xi_l)\right\}
\times
\left[i\gamma^0\ssh A(\xi_l)\right]
\times
\nn
&\times&
\exp\left\{i\gamma^0[\gamma^j k_j-m](\xi_l-y_0)\right\}.
\label{explizit}
\ee

This expression is a uniformly and absolutely converging series
representation for a path-ordered exponential:

\be
G_H(x_0,y_0,\vec k)
=
G_H^0(x_0-y_0,\vec k)
\p\left\{
\int^{x_0}_{y_0}d\xi_0
G_H^0(y_0-\xi_0,\vec k)
[i\gamma^0\ssh A(\xi_0)]
G_H^0(\xi_0-y_0,\vec k)
\right\}.
\nn
\ee

The above derivation is a special case of a more general identity for the
type of path-ordered exponentials encountered here (see appendix A).
The expansion of this formula in powers of $A$ yields:

\be
G_H(x_0,y_0,\vec k)
&=&
G^0_H(x_0-y_0,\vec k)
+
\nn
&+&
\int^{x_0}_{y_0}d\xi_0
G^0_H(x_0-\xi_0,\vec k)
[i\gamma^0\ssh A(\xi_0)]
G^0_H(\xi_0-y_0,\vec k)
+
\nn
&+&
\int^{x_0}_{y_0}d\xi_0\int^{\xi_0}_{y_0}d\eta_0
G^0_H(x_0-\xi_0,\vec k)
[i\gamma^0\ssh A(\xi_0)]
G^0_H(\xi_0-\eta_0,\vec k)
\times
\nn
&&\times
[i\gamma^0\ssh A(\eta_0)]
G^0_H(\eta_0-y_0,\vec k)
+
\nn
&+&
...~.
\label{pertexpa}
\ee

Up to now, only the solution $G_H(x_0,y_0,\vec k)$ of
the homogeneous Dirac equation in the mixed representation has been
investigated. According to the
equation of motion (\ref{dirac}) the inhomogeneous solution 
$iG(x_0,y_0,\vec k)\gamma^0$ must jump by one at $x_0=y_0$. 
The retarded propagator $G_R(x_0,y_0,\vec k)$ vanishes for negative
time differences $x_0-y_0<0$. Due to the 
previous requirement on the argument $g_H(x_0=y_0,y_0,\vec k)=0$ one has for
the homogeneous solution $G_H(x_0=y_0,y_0,\vec k)=1$. Hence, the following
condition has to be fulfilled in order to relate the latter and the retarded
propagator:

\be
iG_R(x_0,y_0,\vec k)\gamma^0
=
\theta(x_0-y_0)
G_H(x_0,y_0,\vec k).
\label{projector}
\ee

It should be noted that if the case of a different, more general coordinate 
with $n^2>0$ should have been investigated at this point, the additional
requirement $n_0>0$ would be needed here in oder to ensure that really the {\it
retarded} propagator is obtained. However, this can always be achieved by a
redefinition of the funcitonal form of the vector potential.

All results obtained for the homogeneous solution of Dirac's equation in the
present mixed representation are linked directly to the Green's function
$G_R(x_0,y_0,\vec k)$ by equation (\ref{projector}). After putting 
equation
(\ref{pertexpa}) into the previous expression, the Heaviside function can be
multiplied to every free homogeneous solution $G_H^0(\zeta_0,\vec k)$:

\be
iG_R(x_0,y_0,\vec k)\gamma^0
&=&
\theta(x_0-y_0)G^0_H(x_0-y_0,\vec k)
+
\nn
&+&
\int^{x_0}_{y_0}d\xi_0
\theta(x_0-\xi_0)G^0_H(x_0-\xi_0,\vec k)
[i\gamma^0\ssh A(\xi_0)]
\times
\nn
&&\times
\theta(\xi_0-y_0)G^0_H(\xi_0-y_0,\vec k)
+
\nn
&+&
\int^{x_0}_{y_0}d\xi_0\int^{\xi_0}_{y_0}d\eta_0
\theta(x_0-\xi_0)G^0_H(x_0-\xi_0,\vec k)
[i\gamma^0\ssh A(\xi_0)]
\times
\nn
&&\times
\theta(\xi_0-\eta_0)G^0_H(\xi_0-\eta_0,\vec k)
[i\gamma^0\ssh A(\eta_0)]
\times
\nn
&&\times
\theta(\eta_0-y_0)G^0_H(\eta_0-y_0,\vec k)
+
\nn
&+&
...~.
\label{projprop}
\ee

This is possible due to the idempotency of the Heaviside function and the
fact that $\theta(x_0-\xi_0)\theta(\xi_0-y_0)=\theta(x_0-y_0)$ if
$\xi_0\in[x_0,y_0]$. Subsequently, in accordance with equation
(\ref{projector}), the result can be reexpressed in terms of free Greens's
functions:

\bmp
\be
iG_R(x_0,y_0,\vec k)
&=&
iG^0_R(x_0-y_0,\vec k)
+
\nn
&+&
\int_{x_0}^{y_0}d\xi_0
iG^0_R(x_0-\xi_0,\vec k)
[i\ssh A(\xi_0)]
iG^0_R(\xi_0-y_0,\vec k)
+
\nn
&+&
\int_{x_0}^{y_0}d\xi_0\int_{\xi_0}^{y_0}d\eta_0
iG^0_R(x_0-\xi_0,\vec k)
[i\ssh A(\xi_0)]
iG^0_R(\xi_0-\eta_0,\vec k)
\times
\nn
&&\times
[i\ssh A(\eta_0)]
iG^0_R(\eta_0-y_0,\vec k)
+
\nn
&+&
...~.
\label{resumt}
\ee
\emp

Note that in the literature slightly different definitions exist for the
propagator which account for the various occurrences of the imaginary unit
$i$.
The full retarded propagator $G_R(x_0,y_0,\vec k)$ inherits the asymptotic
behaviour of the free retarded propagator $G_R^0(x_0-y_0,\vec k)$ by virtue
of the above formula (\ref{resumt}).

\begin{center}
\begin{figure}
\resizebox{!}{6cm}{\includegraphics{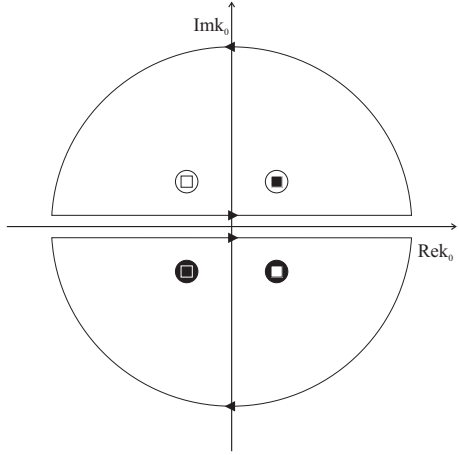}}
\caption{Contour integration in the complex $k_0$-plane for the
determination of the Green's function with the correct assymptotic
behaviour. The circles indicate the pairs of positions to which the poles are 
moved off the real axis by virtue of the corresponding 
$\epsilon$-prescription for the retarded
(black) and the advanced (white) propagator. The squares show
the position of the poles for the Feynman (white) and the reverse Feynman
(black) propagators. For the retarded and the advanced
propagators two poles or no pole is inside a given contour. For the Feynman
and reverse Feynman propagators exactly one pole is always inside any contour.}
\label{contour}
\end{figure}
\end{center}

The full Feynman, i.e. time-ordered propagator cannot be expressed as a
path-ordered exponential because it is defined with mixed boundary
conditions: for the positive energy components at \mbox{$x_0\rightarrow
-\infty$} and for the negative energy components at \mbox{$x_0\rightarrow
+\infty$}. This can also be seen from the free Feynman propagator

\be
iG_F^0(x_0-y_0,\vec k)
&=&
\theta(x_0-y_0)\frac{\gamma^0\omega-\gamma^jk_j+m}{2\omega}
e^{-i\omega(x_0-y_0)}
+
\nn
&+&
\theta(y_0-x_0)\frac{\gamma^0\omega+\gamma^j
k_j-m}{2\omega}e^{+i\omega(x_0-y_0)}
\ee

which is a singular object in this and every mixed
representation. That can be understood by looking at figure 
\ref{contour}.
Thus it is impossible to take its logarithm and express it as an exponential
function. This is why the Feynman propagator cannot be equal to a
path-ordered exponential of the form (\ref{fullt}). 

Figure \ref{contour} shows the contour integrations in the complex
$k_0$-plane which have to be carried out in order to determine the
contributions from the different poles of the corresponding propagator
in momentum representation to that propagator in the mixed representation.
Every pole included inside a contour results in an additive contribution to
the free propagator
proportional to one of the matrices $\ssh k\pm m$. On shell, i.e. for
$k^2=m^2$, these are singular.  The circles in figure \ref{contour} belong
to the retarded (black) and the advanced (white) propagators. It is
important to note that either none of the poles is included in a contour or
both. This means that, if one of these two propagators is non-zero, the two
singular matrices occur in a non-trivial linear combination, which yields an
invertible matrix structure. This is different for the Feynman propagator
and its relatives. There only one pole at a time is included in a contour.
Thus these propagators are always non-invertible.


\subsubsection{Strong-field approximation}

The previous expansion which is appropriate for weak fields 
$A(t)\ll\omega$ could be 
interpreted as an ultraviolet approximation. An infrared expansion requires 
a strong field $A(t)\gg\omega$. It can be obtained by applying the 
resummation formula of 
appendix A in a different way. 
Resumming all scattering
processes with the field for each power of the momentum one obtains:

\be
G_H(x_0,y_0,\vec k)
&=&
\p\left\{i\int^{x_0}_{y_0}d\xi_0\gamma^0\ssh A(\xi_0)\right\}
\times
\nn
&\times&
\p\left[
\int^{x_0}_{y_0}d\xi_0\right.
\p\left\{i\int^{y_0}_{\xi_0}d\theta_0\gamma^0\ssh A(\theta_0)\right\}
\{i\gamma^0[\gamma^jk_j-m]\}
\times
\nn
&&\times
\left.
\p\left\{i\int^{\xi_0}_{y_0}d\theta_0\gamma^0\ssh A(\theta_0)\right\}
\right].
\label{nonpert}
\ee

Now, one could start to expand the outer path-ordered exponential in powers
of the momentum term:

\be
G_H(x_0,y_0,\vec k)
&=&
\p\left\{i\int^{x_0}_{y_0}d\xi_0\gamma^0\ssh A(\xi_0)\right\}
+
\nn
&+&
\int^{x_0}_{y_0}d\xi_0
\p\left\{i\int^{y_0}_{\xi_0}d\theta_0\gamma^0\ssh A(\theta_0)\right\}
\{i\gamma^0[\gamma^jk_j-m]\}
\times
\nn
&&\times
\p\left\{i\int^{\xi_0}_{x_0}d\theta_0\gamma^0\ssh A(\theta_0)\right\}
+
\nn
&+&
\int^{x_0}_{y_0}d\xi_0\int^{\xi_0}_{y_0}d\eta_0
\p\left\{i\int^{y_0}_{\xi_0}d\theta_0\gamma^0\ssh A(\theta_0)\right\}
\{i\gamma^0[\gamma^jk_j-m]\}
\times
\nn
&&\times
\p\left\{i\int^{\xi_0}_{\eta_0}d\theta_0\gamma^0\ssh A(\theta_0)\right\}
\{i\gamma^0[\gamma^jk_j-m]\}
\times
\nn
&&\times
\p\left\{i\int^{\eta_0}_{x_0}d\theta_0\gamma^0\ssh A(\theta_0)\right\}
+
\nn
&+&
...~.
\ee

This corresponds to an expansion in powers of the on-shell energy $\omega$
which can be understood by noting that:
\mbox{$(\gamma^0[\gamma^jk_j-m])^2=\omega^2$}.
The weak-field approximation is based on the investigation of a
given number of the -- otherwise freely propagating -- particle with the
field which could be termed "accelerations".
The strong-field approach comes up to an expansion in powers of what could
be called the
"inertia" because $\omega$ equals the (asymptotic) relativistic mass. In
the lowest order of the strong-field approximation the propagation of a
particle without relativistic mass is governed only by the field. The higher
order terms accord for deviations due to non-vanishing $\omega$.


\subsubsection{Abelian approximation}

All of the above approximations in form of an expansion with respect
to some part of the exponent are based on equation (\ref{resumgen}).
This is different for the Abelian approximation scheme(s), i.e. a commutative
approximation with respect to the Clifford and the charge group algebra.
The lowest order $G_H^{A_0}(x_0,y_0,\vec k)$ of the Abelian approximation is
given by omitting the path-ordering in equation (\ref{fullt}):

\be
G_H^{A_0}(x_0,y_0,\vec k)
=
\exp\left\{
i\gamma^0\int^{x_0}_{y_0}d\xi_0[\gamma^jk_j+\ssh A(\xi_0)-m]
\right\}.
\label{abel0}
\ee

Higher order approximations are not given by additive terms but by splitting
equation (\ref{abel0}) at an ordered set of points: 

\be
G_H^{A_N}(x_0,y_0,\vec k)
=
{\cal P}
\prod_{\nu=0}^N   
\exp\left\{
i\gamma^0\int^{\xi_\nu}_{\xi_{\nu+1}}d\xi_0[\gamma^jk_j+\ssh A(\xi_0)-m]
\right\},
\label{abeln}
\ee

with $x_0=\xi_0<\xi_1<...<\xi_N<\xi_{N+1}=y_0$ for $x_0<y_0$ or 
$x_0=\xi_0>\xi_1>...>\xi_N>\xi_{N+1}=y_0$ for $x_0>y_0$. ${\cal P}$
denotes that the factors are ordered with respect to the index $\nu$ with
the lowest index furthest to the left. The choice of the
intermediate points $\xi_\nu$ is not unique, but in the limit of infinitely
small intervals the result always becomes exact:

\be
&&
\lim_{N\rightarrow\infty}
G_H^{A_N}(x_0,y_0,\vec k)
=
\nn
&=&
\lim_{N\rightarrow\infty}
\prod_{\nu=0}^N   
\exp\left\{
i\gamma^0\int^{\xi_\nu}_{\xi_{\nu+1}}d\xi_0[\gamma^jk_j+\ssh A(\xi_0)-m]
\right\}
=
\nn
&=&
\lim_{N\rightarrow\infty}
\prod_{\nu=0}^N   
\left\{
1+i\gamma^0(\xi_\nu-\xi_{\nu+1})[\gamma^jk_j+\ssh A(\xi_0)-m]
+
{\cal O}\left(\frac{(x_0-y_0)^2}{N^2}\right)
\right\}
=
\nn
&=&
\p\left\{i\gamma^0\int^{x_0}_{y_0}d\xi_0[\gamma^jk_j+\ssh A(\xi_0)-m]\right\}
=
\nn
&=&
G_H(x_0,y_0,\vec k)
\ee

To estimate the error for an interval width of $y_0-x_0=2\Delta$, 
compare the 
lowest order result to the first order where the interval is divided 
into two halves exactly.

\be
\Delta G_H^A
&=&
G_H^{A_0}(0,2\Delta,\vec k)-G_H^{A_1}(0,2\Delta,\vec k)
=
\nn
&=&
G_H^{A_0}(0,2\Delta,\vec k)
-
G_H^{A_0}(0,\Delta,\vec k)G_H^{A_0}(\Delta,2\Delta,\vec k)
\ee 

With the help of the Baker-Campbell-Hausdorff formula: 

\be
\Delta G_H^A
&=&
\exp\left\{
g_H(0,2\Delta,\vec k)
\right\}
-
\exp\left\{
g_H(0,2\Delta,\vec k)
+
[g_H(0,\Delta,\vec k),g_H(\Delta,2\Delta,\vec k)]
+
{\cal O}(\Delta^4)
\right\}
=
\nn
&=&
-[g_H(0,\Delta,\vec k),g_H(\Delta,2\Delta,\vec k)]
+
{\cal O}(\Delta^4)
=
\nn
&=&
-[g_H(0,\Delta,\vec k),g_H(0,2\Delta,\vec k)]
+
{\cal O}(\Delta^4)
=
\nn
&=&
-[g_H(0,\Delta,\vec k),dg_H(0,\Delta,\vec k)/d\Delta]\Delta
+
{\cal O}(\Delta^4).
\label{bch}
\ee

The first occurence of ${\cal O}(\Delta^4)$ results from a Taylor expansion
of secondary and higher commutators. Thus, in leading order of the width of
the interval $\Delta$, the error is proportional to the commutator of the
exponent $g_H$ and its first derivative at an intermediate point of the
interval. For a constant integrand $dg_H(0,\Delta,\vec k)/d\Delta$, i.e. for
a constant gauge field $A$, the Abelian approximation is exact. Higher order
terms are required for fields that lead to a commutator
$[g_H(0,\Delta,\vec k),dg_H(0,\Delta,\vec k)/d\Delta]\Delta$ not small against
$G_H^{A_0}(0,2\Delta,\vec k)$. (This comparison must be based on the
definition of an adequate norm.) 

This result can be compared to the error estimate for the standard form of
expressing a path-ordered exponential as a product of linear factors (see
equation(\ref{poelin})). For that representation one finds:

\be
\Delta G_H
&=&
1+2g_H(0,\Delta,\vec k)
-
[1+g_H(0,\Delta/2,\vec k)][1+g_H(0,3\Delta/2,\vec k)]
+
{\cal O}(\Delta ^2)
=
\nn
&=&
[2g_H(0,\Delta,\vec k)-g_H(0,\Delta/2,\vec k)-g_H(0,3\Delta/2,\vec k)]
-
\nn
&&-
g_H(0,\Delta/2,\vec k)g_H(0,3\Delta/2,\vec k)
+
{\cal O}(\Delta ^2)
=
\nn
&=&
-
g_H(0,\Delta,\vec k)^2
+
[g_H(0,\Delta,\vec k),dg_H(0,\Delta,\vec k)/d\Delta]\Delta/2
+
{\cal O}(\Delta^2).
\ee

Contrary to $\Delta G_H^A$, $\Delta G_H$ does not become zero for a
constant gauge field. In leading order it depends on the actual value of the
exponent $g_H$. Thus, its convergence becomes slow not only for rapid
changes of the gauge field but also for large values of the field and/or
large energies. Even the free propagator then needs
many terms to be approximated sufficiently well.


\subsection{Space-like coordinates}

The general solution scheme for a classical field depending on an arbitrary
rectilinear coordinate $n\cdot x$ leading to equation (\ref{poe}) with the
argument (\ref{argument}) always yields a Green's function
whose boundary conditions are given on a plane normal to $n$. 
Boundary conditions for propagators are given on surfaces with
time-like normal vectors $n^2>0$. Hence, for a field only depending on the 
$x_3$-coordinate ($n^\mu=(0,0,0,1)$), only a Green's function but not a
propagator is given by equations (\ref{poe}) and (\ref{argument}). 

This can also be seen directly. The solution for the argument
$g_H(x_3,y_3;k_0,\vec k_T)$ according to equation (\ref{argument}) is 
given by:

\be
g_H(x_3,y_3;k_0,\vec k_T)
=
-i\gamma^3\int^{x_3}_{y_3}d\xi_3
[\gamma^0k_0+\gamma^Jk_J+\ssh A(\xi_3)-m],
\ee

with an implicit sum over $J\in\{1,2\}$. Repeating the steps that 
led to the free homogeneous
solution of the Dirac equation in the case $n^2>0$ in equation (\ref{freet})
yields:

\be
G_H^0(x_3-y_3;k_0,\vec k_T)
&=&
\gamma^3
\frac{-\gamma^3\sqrt{(k_0)^2-m_T^2}+\gamma^0k_0+\gamma^Jk_J-m}{2\sqrt{(k_0)^2-m_T^2}}
e^{-i\sqrt{(k_0)^2-m_T^2}(x_3-y_3)}
+
\nn
&+&
\gamma^3
\frac{-\gamma^3\sqrt{(k_0)^2-m_T^2}-\gamma^0k_0-\gamma^Jk_J+m}{2\sqrt{(k_0)^2-m_T^2}}
e^{+i\sqrt{(k_0)^2-m_T^2}(x_3-y_3)},
\nn
\ee

with the transverse mass $m_T=\sqrt{|\vec k_T|^2+m^2}$.
This expression, multiplied with $\theta(x_3-y_3)$ in order to obtain a
Green's function from the homogeneous solution, is not proportional to the 
free retarded propagator in this mixed representation.


\subsection{Light-like coordinates}

As mentioned before, the present way to derive a homogeneous solution cannot
be followed if the four-vector $n$ is light-like, because in that case $\ssh
n$ has no inverse.
However, for light-like coordinates there
is a different approach that leads to a solution for $G_H$.
In the case where the normal vector is $n^\mu=(1,0,0,-1)/\sqrt{2}$ equation
(\ref{fourier}) becomes:

\be
\left\{i\gamma_-\frac{d}{dx_-}+\gamma_+[k_-+A_-(x_-)]
-\vec\gamma_T\cdot[\vec k_T+\vec A_T(x_-)]
+\gamma_-A_+(x_-)-m\right\}
~~~~~~~~~~\nn
G_H(x_-,y_-;k_-,\vec k_T)=0,
\label{fourier0}
\ee

with $v_\pm=[v_0\pm v_3]/\sqrt{2}$ and $v_\pm=v^\mp$ where 
$v\in\{\gamma,x,k,A(x_-)\}$. Noting that $\gamma_+\gamma_-/2$ and
$\gamma_-\gamma_+/2$ are two projection operators which project into 
disjoint subspaces of the Clifford algebra and satisfy the completeness 
relation $\gamma_+\gamma_-+\gamma_-\gamma_+=2$ the matrix function 
$G_H(x_-,y_-;k_-,\vec k_T)$ can
be split into
$2G_H(x_-,y_-;k_-,\vec k_T)
=
\gamma_+ G_-(x_-,y_-;k_-,\vec k_T)
+
\gamma_- G_+(x_-,y_-;k_-,\vec k_T)$
with 
$G_\pm(x_-,y_-;k_-,\vec k_T)=\gamma_\pm G_H(x_-,y_-;k_-,\vec k_T)$.
The argument will be suppressed in the following but until the end of the
subsection the above mixed representation is addressed. 
Using this decomposition in equation (\ref{fourier0}) leads to:

\be
i\frac{\gamma_-\gamma_+}{2}\frac{d}{dx_-}G_-
+
\frac{\gamma_+\gamma_-}{2}(k_-+A_-)G_+
-
[\vec\gamma_T\cdot(\vec k_T+\vec A_T)+m]G_H
+
\frac{\gamma_-\gamma_+}{2}A_+G_-
=
0.
\ee

Use has been made of the idempotency of the projectors
$(\gamma_\pm\gamma_\mp/2)^2=\gamma_\pm\gamma_\mp/2$ and their 
projection properties $(\gamma_\pm\gamma_\mp/2)\gamma_\mp=0$ and
$(\gamma_\pm\gamma_\mp/2)\gamma_\pm=\gamma_\pm$.
From here, two equations can be obtained with the help of the projection
operators:

\be
i\frac{d}{dx_-}G_-
+
A_+G_-
-
\frac{1}{2}[\vec\gamma_T\cdot(\vec k_T+\vec A_T)+m]\gamma_-G_+
&=&
0
\nn
(k_-+A_-)G_+
-
\frac{1}{2}[\vec\gamma_T\cdot(\vec k_T+\vec A_T)+m]\gamma_+G_-
&=&
0.
\label{projected}
\ee

The second equation is purely algebraic and can be used to replace $G_+$ in 
the first.

\be
i\frac{d}{dx_-}G_-
+
\frac{1}{2}
\left[\vec\gamma_T\cdot(\vec k_T+\vec A_T)+m\right]
(k_-+A_-)^{-1}
\left[\vec\gamma_T\cdot(\vec k_T+\vec A_T)-m\right]
G_-
=0.
\ee

When postulating a connection between the matrix function $G_-$ and
another $g_-$ in direct analogy to equation (\ref{functional}) the
resulting differential equation is given by:

\be
i\frac{d}{dx_-}g_-
=
-\frac{1}{2}
\left[\vec\gamma_T\cdot(\vec k_T+\vec A_T)+m\right]
(k_-+A_-)^{-1}
\left[\vec\gamma_T\cdot(\vec k_T+\vec A_T)-m\right]
-
A_+
\ee

The equation can be solved by direct integration. As already argued
before, in general the functional $G_-[g_-]$ is given by the
path-ordered exponential of its argument $g_-$. In the present
situation already the absence of non-Abelian charges turns it into an
ordinary exponential because then it only contains the neutral element for
multiplication of the Clifford algebra.
The other component of the matrix function is given by the second of the
equations (\ref{projected}). Finally a homogeneous solution of the
differential equation (\ref{fourier0}) has the form:

\be
&&G_H(x_-,y_-;k_-,\vec k_T)
=
\nn
&=&
\frac{1}{2}\left(
\gamma_+
-
\frac{1}{2}\gamma_-\gamma_+[k_-+A_-(x_-)]^{-1}
\left\{\vec\gamma_T\cdot[\vec k_T+\vec A_T(x_-)]-m\right\}\right)
\times
\nn
&\times&
\p\left[
i\int_{y_-}^{x_-}d\xi_-
\right.
\nn
&&
\left.
\left(\frac{1}{2}
\left\{\vec\gamma_T\cdot[\vec k_T+\vec A_T(\xi_-)]+m\right\}
[k_-+A_-(\xi_-)]^{-1}
\left\{\vec\gamma_T\cdot[\vec k_T+\vec A_T(\xi_-)]-m\right\}
+
A_+(\xi_-)
\right)
\right].
\nn\ee

If one tries to construct a propagator with the help of this homogeneous
solution it can only be retarded or advanced in the light-like
coordinate $x_-$. Alternative approaches can be found in \cite{lightlike}.

In the next chapter the production of fermion-antifermion pairs is described 
based on the results for the fermion propagator in a field that depends on a
time-like coordinate.


\section{Fermion-antifermion pair production}

Here the results for the full propagator in an external field depending on
one rectilinear time-like coordinate are applied to the problem of particle
production due to vacuum polarisation. First it is argued where such a
propagator is of use in describing the physics of fermions in a heavy-ion
collision. Second, a detailed comparison of the different approximation
schemes with the full solution for a given model field is presented.

This calculation can be understood in a twofold way. On the one hand the
field could be really an external field in the sense of the production of
particles via vacuum polarisation. It is determined by the dynamics of the
physical system without taking the back reaction of the particle creation into
account. This field is used to calculate how many particles
would be produced in its presence. This approach is justified if the
process of particle production constitutes merely a small perturbation.
Whether this condition is fulfilled has to be checked afterwards.

On the other hand, the field could already be a self-consistent solution of a
system of equations. For this solution for the classical field one would like 
to know how many particles were created in the process. In the present
scenario such a set of equations would include the Yang-Mills equations with
the expectation value for the current of produced fermions and antifermions
$\left<J_\nu\right>$ and an (initially present) external current
$J_{ext}^\nu$:

\be
\partial^\mu {\cal F}_{\mu\nu}-i[{\cal A}^\mu,{\cal F}_{\mu\nu}]
=
J_{ext}^\nu+\left<J^\nu\right>,
\label{ymc}
\ee

with the gauge field ${\cal A}^\mu$ and the corresponding field tensor
$
{\cal F}_{\mu\nu}
=
\partial_\mu{\cal A}_\nu-\partial_\nu{\cal A}_\mu-i[{\cal A}_\mu,{\cal A}_\nu]
$
in the adjoint representation.
The expectation value for the current can be obtained from
the causal propagator \cite{sk}:

\be
\left<J^\nu\right>\sim{\rm tr}\left\{\gamma^\nu G_C(x,x)\right\},
\label{evc}
\ee 

where the trace is only running over the matrices of the Clifford algebra
and with the definition:
$
G_C(x,x)
=\lim_{\epsilon\rightarrow 0}[G_C(x+n\epsilon,x)+G_C(x,x+n\epsilon)]/2
$
with $n^2>0$.
Higher-order radiative corrections are suppressed by powers of the
coupling constant which are not compensated by powers of the
classical field.
The causal propagator can be reexpressed as a linear combination of the 
retarded, the advanced, and the on-shell propagator:

\be
G_C(x,y)=\frac{1}{2}[G_R(x,y)+G_A(x,y)+G_S(x,y)].
\ee

The advanced propagator can be obtained from the homogeneous solution
$G_H(x_0,y_0,\vec k)$ for the equation of motion for the Dirac Green's
function by the relation:

\be
i G_A(x_0,y_0,\vec k)\gamma^0
=
-\theta(y_0-x_0)G_H(x_0,y_0,\vec k).
\ee

The on-shell propagator can be reexpressed in terms of the retarded and
advanced one-particle scattering operators \cite{baltz}:

\be
G_S(x,y)
=
\int\frac{d^4k}{(2\pi)^4}
2\pi\delta(k^2-m^2)
G_R^0(q){\cal T}_R(q,k)(\ssh k+m){\cal T}_A(k,p)G_A^0(p).
\ee      

The scattering operators are defined according to equation (\ref{opso})
below. In the present framework all required propagators are known as
functionals of the classical gauge fields. Hence, equation (\ref{ymc}) with
equation (\ref{evc})
constitutes an integro-differential equation for the classical gauge field.
Its solution would yield the form of the field. The expectation value for
the produced fermions and antifermions could be calculated from this field.

\begin{center}
\begin{figure}
\resizebox{!}{6cm}{\includegraphics{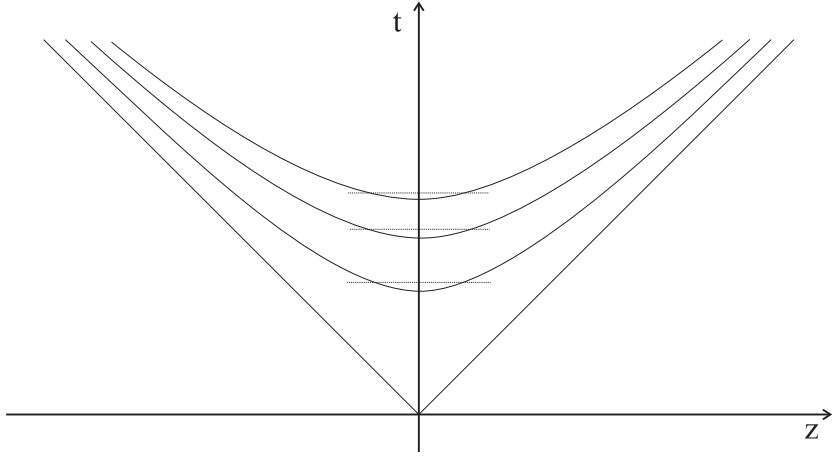}}
\caption{Time dependence as approximation to the situation found in the
central region of a boost-invariant system. The proper time $\tau$ is
constant on the hyperbolas.}
\label{lightcone}
\end{figure}
\end{center}

Let us consider a model for the classical radiation-field in an
ultra-relativistic heavy-ion collision. According to Bjorken 
\cite{bjorken}, the
mid-rapidity region in a heavy-ion collision is characterised by
boost-invariant quantities, i.e. boost-invariant along the beam direction.
Let us consider a central collision in a symmetric system in the
center-of-mass frame. For absolute values of the longitudinal coordinate
$|z|$ smaller than the kinematic time $t$ in this frame of reference the
dependence on proper time $\tau=\sqrt{t^2-z^2}$ is approximated well by a
dependence on the kinematic time $t$ (see figure
\ref{lightcone}). Most of the energy is deposited during $t<t_{in}$ close to 
the collision point at $z=0$ \cite{initial}.
Hence, in good
approximation, for $t>t_{in}$ and $|z|<t_{in}$, an in general proper-time
$\tau$ dependent energy density can be reexpressed as an energy density
depending on kinematic time $t$. Let the entire energy density be initially
stored in an electric field of the form $E_\eta(\tau)$ (component along the
hyperbolas). In the present approach, this
fact is consistently approximated by a storage of the energy in
$E_z(t)$. In temporal gauge ($A_0=0$) or even in Lorentz gauge
this is equivalent to a gauge field $A_z(t)$. For this form of gauge field,
the retarded propagator has been derived in the previous section.

In order to proceed, one needs to know how to describe particle production
based on a given propagator. With the Fourier transform of the retarded 
propagator:

\be
G_R(x,y)
=
\int\frac{d^4p}{(2\pi)^4}\frac{d^4q}{(2\pi)^4}
e^{-iq\cdot x}e^{+ip\cdot y}
G_R(q,p)
\ee

one can write the implicit definition of the corresponding one-particle
scattering operator $\cal T$ in momentum space as:

\be
G_R(q,p)
=
(2\pi)^4\delta^{(4)}(q-p)G_R^0(p)
+
G_R^0(q)\times{{\cal T}}(q,p)\times{G}_R^0(p).
\ee

Explicitly, it is given by:

\be
{\cal T}(q,p)
=
\ssh A(q-p)
+
\int\frac{d^4k}{(2\pi)^4}\frac{d^4l}{(2\pi)^4}
\ssh A(q-k)
G_R(k,l)
\ssh A(l-p).
\label{opso}
\ee

In the following ${\cal T}$ always denotes the {\it retarded}  one-particle
scattering operator.
The Born approximation, the expansion to lowest order in the fields, is
given by:

\be
{{\cal T}}^B(q,p)=\ssh{A}(q-p).
\label{born}
\ee

This term is always contained in the scattering operator. The different
approximation schemes discussed in the last section lead to differences in
the remaining non-Bornian part in equation (\ref{opso}). In the presence of
a purely time-dependent field the retarded one-particle scattering operator
becomes:

\be
{{\cal T}}(q,p)
&=&
(2\pi)^3\delta^{(3)}(\vec q-\vec p)
\times
\nn
&&\times
\left[
\ssh A(q_0-p_0)
+
\int dx_0dy_0e^{+iq_0x_0}e^{-ip_0y_0}
\ssh A(x_0) G_R(x_0,y_0,\vec p) \ssh A(y_0)
\right]
=
\nn
&=&
(2\pi)^3\delta^{(3)}(\vec q-\vec p)
{{\cal T}}(q_0,p_0)
=
(2\pi)^3\delta^{(3)}(\vec q-\vec p)
\left[
{{\cal T}^B}(q_0,p_0)
+
{{\cal T}^{NB}}(q_0,p_0)
\right],
\ee

with the non-Bornian part ${{\cal T}^{NB}}(q_0,p_0)$. 
Due to the occurence of the Dirac $\delta$-distribution the
conservation of the total three-momentum becomes obvious. In purely
time-dependent fields the fermion-antifermion pairs are always produced in a 
{\it back-to-back} configuration.
Terms of higher order 
in the gauge field -- for instance in the weak-field expansion -- can be
obtained by replacing the full propagator in the scattering operator by terms
from equation (\ref{resumt}). Analogously, replacing the full
propagator by various approximations leads to the corresponding
approximations for the one-particle scattering operator.

From the retarded one-particle scattering operator ${\cal T}_R$ the
expectation value of produced pairs can be obtained:

\be
\left<n\right>
=
\int 
\frac{d^3q}{2(2\pi)^3q_0} 
\frac{d^3p}{2(2\pi)^3p_0}
\left|\bar{u}(q){{\cal T}}_R(q,-p)v(p)\right|^2,
\ee

where a summation over the spin degrees of freedom of the unit spinors
$\bar{u}(q)$ and $v(p)$ is understood and where $p_0=\sqrt{|\vec p|^2+m^2}$ 
and $q_0=\sqrt{|\vec q|^2+m^2}$.
For the special form of the scattering operator in spatially homogeneous 
purely time-dependent situations this simplifies to:

\be
\left<n\right>
=
\frac{V}{4(2\pi)^3}\int 
\frac{d^3p}{{p_0}^2} 
\left|\bar{u}(p_0,-\vec p){\cal T}(p_0,-p_0)v(p_0,+\vec p)\right|^2,
\ee

where use has been made of the relation 
$[\delta^{(3)}(\vec p-\vec q)]^2=V\delta^{(3)}(\vec p-\vec q)/(2\pi)^3$. 
Carrying out the spin summation leads to:

\be
\frac{4(2\pi)^3}{V}\frac{d\left<n\right>}{d^3p}
=
{\rm tr}\left\{
{\cal T}(p_0,-p_0)
\frac{\gamma^0p_0+\gamma^jp_j-m}{p_0}
\gamma^0
{\cal T}^\dagger(p_0,-p_0)
\gamma^0
\frac{\gamma^0p_0-\gamma^jp_j+m}{p_0}
\right\}.
\label{momentumspectrum}
\ee

In order to gain some insight into the behaviour of the differential
expectation value (or momentum spectrum) and some information on the quality 
of the different
approximations without having to solve the Yang-Mills equations (\ref{ymc})
beforehand, the different formulae are going to be evaluated for a special
choice of the field:

\be
A_\mu(t)=g_{3\mu}A_{in}e^{-t/t_0}\theta(t).
\label{specfield}
\ee

Many other forms could have been taken. This choice was also inspired by a
numerical study \cite{bhaleraonayak} which indicates that the field decays
in a similar fashion. In any case, the actual form of the
classical field has to be determined in a self-consistent calculation.

For this field, the one-particle scattering operator in Born approximation
(\ref{born}) is given by:

\be
{\cal T}^B(2\omega)=\frac{\gamma^3 A_{in}t_0}{1+2it_0\omega}.
\label{bornspec}
\ee

The following subchapters show the various approximations of the remaining 
part of the retarded one-particle scattering operator in the field
(\ref{specfield}).


\subsubsection{The weak-field approximation}

The lowest-order weak-field term of the full propagator is given by the
free propagator (equation (\ref{projector}) together with equation 
(\ref{freet})). Here, the non-Bornian part of the one-particle scattering 
operator is:

\be
{\cal T}^{UV}
=
-i
\gamma^3\frac{\gamma^0\omega+\gamma^jk_j-m}{2\omega}\gamma^3
{\cal T}^{UV}_+
-i
\gamma^3\frac{\gamma^0\omega-\gamma^jk_j+m}{2\omega}\gamma^3
{\cal T}^{UV}_-,
\label{uvspec}
\ee

with

\be
{\cal T}^{UV}_{\pm}
=
A_{in}^2\int_0^\infty dx_0\int_0^{x_0}dy_0
e^{i\omega(x_0+y_0)}e^{\pm i\omega(x_0-y_0)}e^{-x_0/t_0}e^{-y_0/t_0}
=
\frac{(A_{in}t_0)^2/2}{[1-i\omega t_0][1-i(\omega\pm\omega)t_0]}.
\nn\ee


\subsubsection{The strong-field approximation}

For any purely time-dependent field, the general expression for the
homogeneous solution in the lowest-order strong-field approximation:

\be
G_H^{IR}(x_0,y_0,\vec k)
=
\p\left\{i\int_{y_0}^{x_0}d\xi_0\gamma^0\ssh A(\xi_0)\right\}
\ee

is not much simpler to evaluate 
than the exact solution. However, for a field of constant direction 
$A_\mu(t)=A_\mu\times f(t)$ the path-ordering can be dropped. 
In the lowest-order strong-field approximation the non-Bornian part of the
one-particle scattering operator is:

\be
{\cal T}^{IR}
=
-i
\frac{\gamma^0-\gamma^3}{2}
{\cal T}^{IR}_+
-i
\frac{\gamma^0+\gamma^3}{2}
{\cal T}^{IR}_-,
\label{irspec}
\ee

with:

\be
{\cal T}^{IR}_{\pm}
&=&
{A_{in}}^2
\int_0^\infty dx_0\int_0^{x_0}dy_0
e^{i\omega(x_0+y_0)}e^{-x_0/t_0}e^{-y_0/t_0}
\exp\left\{\mp iA_{in}t_0[e^{-x_0/t_0}-e^{-y_0/t_0}]\right\}
=
\nn
&=&
(A_{in}t_0)^2
\sum_{\mu=0}^\infty
\frac{(\mp iA_{in}t_0)^\mu}{\mu!}
\frac{1}{\mu+1-i\omega t_0}
\sum_{\nu=0}^{\infty}
\frac{(\pm iA_{in}t_0)^\nu}{\nu!}
\frac{1}{\nu+\mu+2-2i\omega t_0},
\ee

where use has been made of the uniform convergence of the exponential series
for bounded arguments.
With formula 6.5.29 in \cite{as}:

\be
\gamma^*(a,z)
=
\frac{1}{\Gamma(a)}
\sum_{n=0}^{\infty}\frac{(-z)^n}{(a+n)n!}
\ee

for a bounded norm of $A_{in}t_0$:

\be
{\cal T}_{\pm}^{IR}
=
(A_{in}t_0)^2
\sum_{\mu=0}^\infty
\frac{(\mp iA_{in}t_0)^\mu}{\mu!}
\frac{1}{\mu+1-i\omega t_0}
\gamma^*(\mu+2-2i\omega t_0,\mp iA_{in}t_0)
\Gamma(\mu+2-2i\omega t_0).
\nn
\ee

Formula 6.5.4 in \cite{as}

\be
\gamma^*(a,z)
=
\frac{z^{-a}}{\Gamma(a)}\gamma(a,z)
\ee

leads to:

\be
{\cal T}_{\pm}^{IR}
&=&
-
(\mp A_{in}t_0)^{2i\omega t_0}
\sum_{\mu=1}^\infty
\frac{1}{\mu!}
\frac{\gamma(\mu+1-2i\omega t_0,\mp iA_{in}t_0)}{1-i\omega t_0/\mu}.
\ee

In the case of multiple charges, $A_{in}$ can be decomposed according to:
$A_{in}=A^a_{in}T^a$ where the $T^a$ are the generators of the corresponding
algebra. Due to the requirement of unitarity these generators have to be
hermitian. This is also true for any linear combination of the generators with 
real coefficients. Thus every matrix $A^a_{in}T^a$ with real $A^a_{in}$ can
be diagonalised, yielding:

\be
A^a_{in}T^a=\sum_{n=1}^{N}\lambda_n\left|n\right>\left<n\right|,
\ee

with the eigenvalues $\lambda_n$ and the $N$ orthonormal eigenvectors
$\left|n\right>$. The $\left|n\right>\left<n\right|$ are projectors onto
subspaces of different charges. Thus one gets:

\be
{\cal T}_{\pm}^{IR}
=
-
\sum_{n=1}^{N}
\left|n\right>\left<n\right|
(\mp \lambda_n t_0)^{2i\omega t_0}
\sum_{\mu=1}^\infty
\frac{1}{\mu!}
\frac{\gamma(\mu+1-2i\omega t_0,\mp i\lambda_n t_0)}{1-i\omega t_0/\mu}.
\ee


\subsubsection{The modified strong-field approximation}

In the present situation the special form of the field allows for a variation 
of the strong-field
approximation, where the component of the momentum parallel to the field --
in this case $k_3$ -- is included in the exponent of the lowest-order
expression $A_3\rightarrow A_3+k_3$:

\be
G_H^{IR'}(x_0,y_0,\vec k)
=
\p\left\{i\int_{y_0}^{x_0}d\xi_0\gamma^0\gamma^3 [k_3+A_3(\xi_0)]\right\}.
\ee

In the lowest-order modified strong-field approximation the additional part of
the one-particle scattering operator beyond the Born approximation is:

\be
{\cal T}^{IR^\prime}
=
-i
\frac{\gamma^0-\gamma^3}{2}
{\cal T}^{IR^\prime}_+
-i
\frac{\gamma^0+\gamma^3}{2}
{\cal T}^{IR^\prime}_-,
\label{enhirspec}
\ee

with:

\be
{\cal T}^{IR^\prime}_{\pm}
&=&
{A_{in}}^2
\int_0^\infty dx_0\int_0^{x_0}dy_0
e^{i(\omega\pm k_3)(x_0+y_0)}e^{-x_0/t_0}e^{-y_0/t_0}
\exp\left\{\mp iA_{in}t_0[e^{-x_0/t_0}-e^{-y_0/t_0}]\right\}.
\nn
\ee

Repeating the above steps leads to:

\be
{\cal T}_{\pm}^{IR^\prime}
&=&
-
\sum_{n=1}^{N}
\left|n\right>\left<n\right|
(\mp\lambda_n t_0)^{2i\omega t_0}
\sum_{\mu=1}^\infty
\frac{1}{\mu!}
\frac
{\gamma(\mu+1-2i\omega t_0,\mp i\lambda_n t_0)}
{1-i(\omega\pm k_3)t_0/\mu}.
\ee


\subsubsection{The Abelian approximation}

In the lowest-order Abelian approximation the interacting part of the
retarded propagator is:

\be
{\cal T}^{A}
=
+i
{\cal T}^{A}_+
+i
{\cal T}^{A}_-,
\label{aspec}
\ee

with

\be
{\cal T}^{A}_{\pm}
&=&
{A_{in}}^2
\int_0^\infty dx_0\int_0^{x_0}dy_0
e^{i\omega(x_0+y_0)}e^{\pm i\Omega(x_0-y_0)}e^{-x_0/t_0}e^{-y_0/t_0}
\frac{\gamma^0\Omega\pm[\gamma^Jk_J-\gamma^3K_3+m]}{2\Omega},
\nn\ee

with the generalised energy
$\Omega=\sqrt{{m_T}^2+{K_3}^2}$
and the generalised momentum 
\mbox{$K_3=k_3+A_{in}t_0(e^{-x_0/t_0}-e^{-y_0/t_0})/(x_0-y_0)$}.

Decomposition with respect to multiple charges leads to:

\be
{\cal T}^{A}_{\pm}
&=&
\sum_{n=1}^{N}
\left|n\right>\left<n\right|
{\lambda_n}^2
\times
\nn
&&\times
\int_0^\infty dx_0\int_0^{x_0}dy_0
e^{i\omega(x_0+y_0)}e^{\pm i\Omega_n(x_0-y_0)}e^{-x_0/t_0}e^{-y_0/t_0}
\frac{\gamma^0\Omega_n\pm[\gamma^Jk_J-\gamma^3(K_3)_n+m]}{2\Omega_n}
.\nn
\ee

with the generalised energy
$\Omega_n=\sqrt{{m_T}^2+{(K_3)_n}^2}$
and the generalised momentum 
$(K_3)_n=k_3+\lambda_n t_0(e^{-x_0/t_0}-e^{-y_0/t_0})/(x_0-y_0)$
belonging to the respective eigenvalue $\lambda_n$.  The modification to the
longitudinal momentum is equal to the arithmetic average of the gauge field
over the interval $[x_0,y_0]$.  Hence the Abelian approximation can be
interpreted as the description of the propagation of the fermions with their
arithmetically averaged canonical momentum. In the weak-field expansion
they are propagated with their asymptotic kinematic momentum. Higher orders
in the Abelian approximation scheme make better approximations similar to a
Fourier series.  The particle is propagated with its canonical momentum
averaged over every piece of the trajectory. The finer the partitioning of
the path, the closer the average canonical momentum is to its actual
value in a particular interval. In the weak-field perturbative
approximation scheme the particle is always propagated with its asymptotic
kinetic momentum. For higher orders it only interacts with the field more
and more often.

For the strong-field approximation and the modified strong-field approximation the
replacement of the path-ordered exponential by an exponential function is only
possible due to the special form of the field. In the Abelian approximation
scheme this exchange is possible in the presence of an arbitrary field.

The decomposition with respect to the charge projectors is also possible for
the full non-Bornian part and
the omnipresent Born part of the one-particle scattering operator
${\cal T}^B=\sum_{n=1}^N\left|n\right>\left<n\right|{\cal T}^B_n$. 
Hence, the whole operator can always be
decomposed with respect to the same projectors
${\cal T}=\sum_{n=1}^N\left|n\right>\left<n\right|
[{\cal T}^B_n+{\cal T}^{NB}_n]$.  In the squared expression
needed to calculate the expectation value the contributions belonging to
the different projectors do not mix. They lead to a sum over the
expectation values for the different charge sub-spaces
${\rm tr}_c
\left\{
\sum_{n'=1}^N\left|n'\right>\left<n'\right|{\cal T}_{n'} 
\sum_{n"=1}^N\left|n"\right>\left<n"\right|{\cal T}_{n"}^*
\right\}
=
\sum_{n'=1}^N\left|{\cal T}_{n'}\right|^2
{\rm tr}_c\left\{\left|n'\right>\left<n'\right|\right\}
=
\sum_{n'=1}^N\left|{\cal T}_{n'}\right|^2$,
where ${\rm tr}_c$ denotes the trace over the generators of the
charge group. If the eigenvectors $\left|n'\right>$ are normalised, the
remaining trace is equal to unity. Due to these facts, it suffices to
compare the contributions from the different approximation schemes for one of
the sub-spaces.

It is always possible to measure all momenta, energies, and gauge field
strengths in units of a scale parameter with the dimension of momentum. Then 
all lengths and times have to be given in units of inverse momenta. In the
following, the eigenvalue belonging to the corresponding subspace is chosen
as scale parameter and is going to be called $A_{in}/g$ again. The 
calculations 
are carried out assuming that all the energy of the system is
included in one of the sub-spaces.

The expected energy density produced in a central heavy-ion collision at LHC
(Pb-Pb at $\sqrt{s}$= 5.5 TeV) is $\epsilon\approx$ 1000 GeV/fm$^3$
\cite{initial,epsilon,alpha}. For the strong coupling constant one expects:
$\alpha_s \approx$ 0.15 \cite{alpha}.
If all the energy density was deposited in the field sector a rough estimate
for the initial gauge field magnitude would be 
$A_{in}\approx\sqrt{g\sqrt{2 \epsilon}}\approx 2{\rm GeV}$.
For RHIC (Au-Au at $\sqrt{s}$= 200 GeV) the typical coupling constant is
around $\alpha_s\approx$ 0.33 and the initial energy density
$\epsilon\approx$ 50 GeV/fm$^3$. This would lead to $A_{in}\approx 1{\rm
GeV}$. With decay times in the range from 0.1fm/c to 0.5fm/c this leads to
$A_{in}t_0$ between 0.5 and 5.0. Here only massless particles are
investigated.

The expressions for the Born (\ref{bornspec}) and the weak-field
approximation (\ref{uvspec}) can be evaluated straightforwardly. For the
strong-field (\ref{irspec}) and the modified strong-field approach (\ref{enhirspec})
the few first terms of the infinite series representations suffice for
obtaining an accurate result. The integrals for the Abelian approximation
have to be treated with standard numerical methods. The exact solution
requires the handling of path-ordered exponentials and subsequent
integrations. 

\begin{center}
\begin{figure}
\resizebox{!}{12cm}{\includegraphics{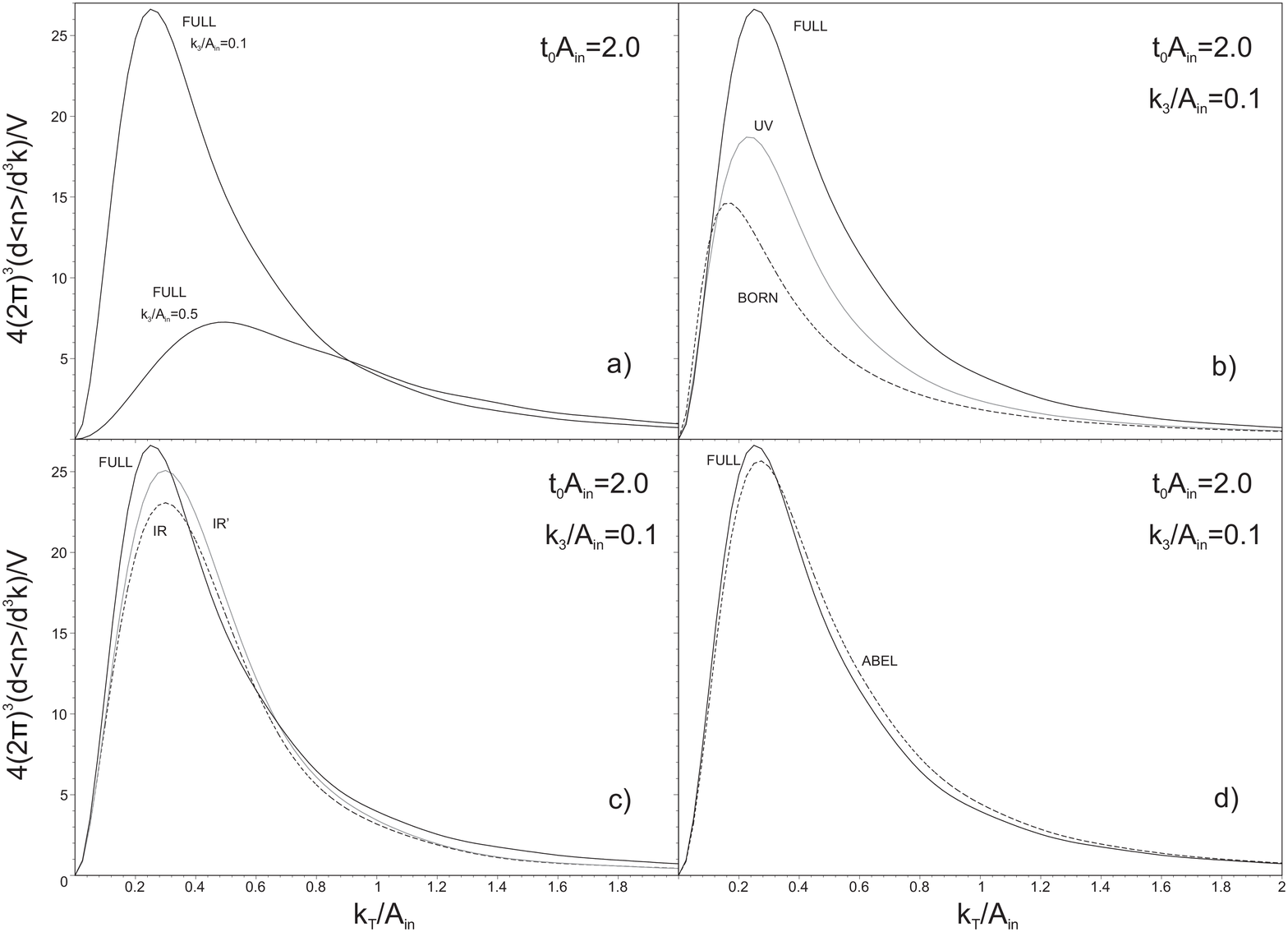}}
\caption{
Momentum spectrum of produced massless 
fermion-antifermion pairs versus  
transverse momentum and with
the decay time $t_0A_{in}=2.0$: a) Exact result for different 
values of the
longitudinal momentum. In plots b), c), and d) the longitudinal momentum is
fixed at $k_3=0.1A_{in}$ b) Exact result (solid) compared to the Born (dashed)
and the weak-field (grey) approximation. c) Exact result (solid) compared to
the strong-field (dashed) and the modified strong-field (grey) approximation. d)
Exact result (solid) compared to the Abelian (dashed) approximation.
}
\label{approximations}
\end{figure}
\end{center}

\begin{center}
\begin{figure}
\resizebox{!}{6cm}{\includegraphics{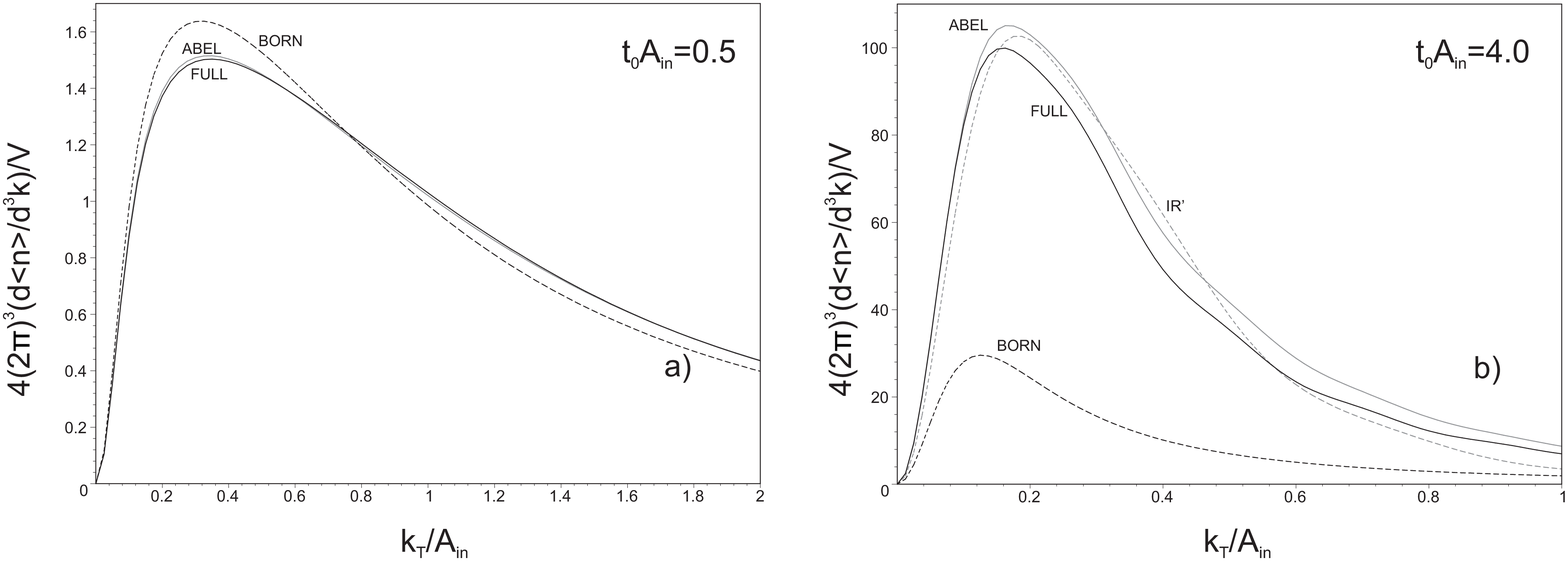}}
\caption{
Momentum spectrum (solid) of produced massless 
fermion-antifermion pairs versus  
transverse momentum compared to the Born (dashed) and the Abelian (grey) 
approximation for fixed
longitudinal momentum $k_3=0.1A_{in}$ and for different values of the
decay time: a) $A_{in}t_0=0.5$ and b)
$A_{in}t_0=4.0$. In plot b) the enhanced strong-field approximation
(dashed grey) is shown, too. 
}
\label{timescales}
\end{figure}
\end{center}

The general aspects of the exact solution for the momentum spectrum
(\ref{momentumspectrum}) are best seen in figures (\ref{approximations}a)
and (\ref{timescales}). As a function of the transverse momentum $k_T$ it
peaks once and shows no further relative extrema or other distinct
structures. For increasing values of the parameter $A_{in}t_0$ (from figure
(\ref{timescales}a) over figure (\ref{approximations}a) to figure
(\ref{timescales}b)) the peak in the transverse momentum spectrum becomes more
pronounced for a fixed value of the longitudinal momentum $k_3$. In other
words it increases in height and decreases in width (see especially the
different scale of the transverse momentum axis in figure
(\ref{timescales}b)). Actually, the differential expectation value is a
function of the variable $\omega t_0$. Hence the width of the
transverse-momentum spectrum for massless particles at mid-rapidity scales 
exactly inversely proportionally to $t_0$. The same holds still after $\omega$
and $t_0$ have been rescaled with $A_{in}$.
For fields of a functional form analogous to that of
the present special model field the peak height seems to
be strictly monotonically decreasing with increasing longitudinal
momentum, as is suggested in figure (\ref{approximations}a). Further, at
zero transverse momentum no particles are produced; the fermions and
antifermions are never produced with momenta along the direction of the field
but preferentially with momenta perpendicular to the field.

A comparison of the different approaches shows that for large momenta all
approximations and the exact solution tend towards the Born result. This is
due to the form of the one-particle scattering operator (\ref{opso}).
Together with the Born approximation all other graphs tend toward zero for
higher particle energies. As shown in figure (\ref{approximations}b) the
Born approach overestimates the exact value for low momenta but
underestimates it for high momenta. The weak-field approximation is an
improvement compared to the Born approach for most values of the transverse
momentum. Looking at figure (\ref{approximations}c) the strong-field and the
modified strong-field are generally closer to the exact result than the
weak-field approximation. However, for more general forms of time-dependent
fields the propagators in these schemes are not much simpler to deal with
than the full one. The modified strong-field approximation even ceases to be
available because the terms longitudinal and transverse might no longer be
well-defined with respect to the field. For all momenta the Abelian
approximation scheme (see figure (\ref{approximations}d)) is closest to the
exact values. The largest deviations are found
for small energies and large values of the parameter $A_{in}t_0$ (compare
figures (\ref{approximations}d), (\ref{timescales}a), and
(\ref{timescales}b)). The reason is that there the situation is maximally
non-Abelian, i.e. there the condition that the typical commutator of
the exponent $g_H(x_0,y_0,\vec k)$ at different points is negligible with
respect to the typical propagator is least well satisfied (see also equation
(\ref{bch})). While for low values of $A_{in}t_0$ the Born approximation 
is reasonably good it is not appropriate for large values (see figure
(\ref{timescales})).


\section{Summary}

The exact homogeneous solutions for the Dirac equation in a gauge field
depending on one rectilinear coordinate has been presented. An alternative
way had to be taken for a dependence on a light-like coordinate. In the case
where this coordinate was time-like, the retarded propagator has been
constructed from the homogeneous solution. The analogous result for a
space-like coordinate was seen to constitute a Dirac Green's function but
not a propagator.

For the situation of a time-like coordinate various approximation schemes
for the exact solution have been determined. Explicitly, these are the
weak-field approximation, the strong-field approach, and the Abelian
approximation. Additionally, a larger variety of approximations can be
obtained with the help of the general resummation formula.

Subsequently, the retarded fermion propagator and all the lowest orders of
the various approximation schemes in the presence of a gauge field depending
on one rectilinear time-like coordinate have been used to calculate the
momentum spectrum of produced fermion-antifermion pairs. The
resulting expressions are evaluated for a decaying model field and the
results are mutually compared for parameters expected to be found in
ultra-relativistic heavy-ion collisions. In this situation an additional
modified strong-field approximation could be obtained.

In the present situation, the exact momentum spectrum is a singly peaked
function of the transverse momentum with no further distinct
structure. The quality of the approximations increases from the Born
approach over the lowest-order weak-field, strong-field, modified
strong-field, towards the Abelian approximation. It should be mentioned that in more
general situations the strong-field and the modified strong-field approach are
not much simpler to evaluate than the full result. The model parameter is
$A_{in}t_0$. It is the product of the initial magnitude of the gauge field
$A_{in}$ and the decay time scale of the field. For the smallest expected
values the Born approximation is still acceptable. Nevertheless, the
other schemes like the Abelian or the enhanced strong-field are even better.
For the highest values of the decay time only the latter come close to the 
exact result. Hence, in order to ensure the maximum possible independence
from the scale parameter $A_{in}t_0$ without having to evaluate the exact
solution it would be best to use the Abelian approximation for
self-consistent calculations.


\section*{Acknowledgements}

Helpful discussions with A.~D\"oring, S.~Hofmann, K.~Kajantie, A.~Mishra, 
J.~Reinhardt,
D.~Rischke, J.~Ruppert, and S.~Schramm are gratefully acknowledged. In this
context, I am particularly indebted to J.~Reinhardt and D.~Rischke for their
effort to help me to complete this work. It has been financially
supported by the Graduiertenf\"orderung des Landes Hessen and by the
Gesellschaft f\"ur Schwerionenforschung.



\appendix

\section{General resummation formula}

In general, a path-ordered exponential with an integrand depending on a
single variable $\xi_0$ can be rewritten in the following way:

\bmp\be
\p\left\{\int^{x_0}_{y_0}d\xi_0[B(\xi_0)+C(\xi_0)]\right\}
&=&
\sum_{l=0}^\infty
\int^{x_0}_{y_0}d\xi_1\int^{\xi_1}_{y_0}d\xi_2...\int^{\xi_{l-1}}_{y_0}d\xi_l
\nn
&~&
\p\left\{\int_{\xi_1}^{x_0}d\xi_0 B(\xi_0)\right\}
C(\xi_1)
\times
\nn
&\times&
\p\left\{\int_{\xi_2}^{\xi_1}d\xi_0B(\xi_0)\right\}
C(\xi_2)
\times
\nn
&\times&
...
\times
\nn
&\times&
\p\left\{\int_{\xi_l}^{\xi_{l-1}}d\xi_0 B(\xi_0)\right\}
C(\xi_l)
\times
\nn
&\times&
\p\left\{\int_{y_0}^{\xi_l}d\xi_0 B(\xi_0)\right\}.
\ee\emp

Making use of the group property valid for the present path-ordered 
exponentials:

\be
\p\left\{\int_{z_0}^{x_0}d\xi_0 B(\xi_0)\right\}
\times
\p\left\{\int_{y_0}^{z_0}d\xi_0 B(\xi_0)\right\}
=
\p\left\{\int_{y_0}^{x_0}d\xi_0 B(\xi_0)\right\}
\label{addth}
\ee

the above equation can be reexpressed as:

\be
&\p&\left\{\int^{x_0}_{y_0}d\xi_0[B(\xi_0)+C(\xi_0)]\right\}
=
\nn\nn
=
&\p&\left\{\int^{x_0}_{y_0}d\xi_0B(\xi_0)\right\}
\sum_{l=0}^\infty
\int^{x_0}_{y_0}d\xi_1\int^{\xi_1}_{y_0}d\xi_2...\int^{\xi_{l-1}}_{y_0}d\xi_l
\times
\nn
\times
&\p&\left\{\int_{x_0}^{y_0}d\xi_0 B(\xi_0)\right\}
\p\left\{\int_{\xi_1}^{x_0}d\xi_0 B(\xi_0)\right\}
C(\xi_1)
\p\left\{\int^{\xi_1}_{y_0}d\xi_0 B(\xi_0)\right\}
\times
\nn
\times
&\p&\left\{\int_{x_0}^{y_0}d\xi_0 B(\xi_0)\right\}
\p\left\{\int_{\xi_2}^{x_0}d\xi_0 B(\xi_0)\right\}
C(\xi_2)
\p\left\{\int^{\xi_2}_{y_0}d\xi_0 B(\xi_0)\right\}
\times
...
\times
\nn
\times
&\p&\left\{\int_{x_0}^{y_0}d\xi_0 B(\xi_0)\right\}
\p\left\{\int_{\xi_l}^{x_0}d\xi_0 B(\xi_0)\right\}
C(\xi_l)
\p\left\{\int^{\xi_l}_{y_0}d\xi_0 B(\xi_0)\right\}
=
\nn\nn
=
&\p&\left\{\int^{x_0}_{y_0}d\xi_0 B(\xi_0)\right\}
\times
\nn
&&\times
\p\left[
\p\left\{\int_{x_0}^{y_0}dz_0 B(z_0)\right\}\right.
\times
\nn
&&~~~~~\times
\left.\int_{x_0}^{y_0}d\xi_0
\p\left\{\int_{\xi_0}^{x_0}dz_0 B(z_0)\right\}
C(\xi_0)
\p\left\{\int^{\xi_0}_{y_0}dz_0 B(z_0)\right\}
\right].
\nn
\label{general}
\ee

Again, by virtue of the group property the most compact form is given by:

\be
&\p&\left\{\int^{x_0}_{y_0}d\xi_0[B(\xi_0)+C(\xi_0)]\right\}
=
\nn
=
&\p&\left\{\int^{x_0}_{y_0}d\xi_0 B(\xi_0)\right\}
\times
\nn
\times
&\p&\left[
\int^{x_0}_{y_0}d\xi_0
\p\left\{\int_{\xi_0}^{y_0}dz_0 B(z_0)\right\}
C(\xi_0)
\p\left\{\int^{\xi_0}_{y_0}dz_0 B(z_0)\right\}
\right].
\label{resumgen}
\ee

Summarising, 
more general resummations are possible in which the dominant quantity can be
chosen arbitrarily. Further, the above steps can be repeated so as to 
resum the obtained quantity several times, e.g. after splitting $C(\xi_0)$
into a dominant part and a deviation.

The above result can also be obtained in another way. Write the Fourier
transformed Dirac equation in a generic form, where the dependence
on $\vec k$ will not be denoted in the following:

\be
\left[\frac{d}{dx_0}-B(x_0)-C(x_0)\right] G_H(x_0,y_0)=0.
\label{genericdirac}
\ee

With the ansatz

\be
G_H(x_0,y_0)=U(x_0,y_0)\hat G_H(x_0,y_0)
\ee

and the product rule for differentiation one obtains:

\be
\left[\frac{d}{dx_0}U(x_0,y_0)\right]\hat G_H(x_0,y_0)
+
U(x_0,y_0)\left[\frac{d}{dx_0}\hat G_H(x_0,y_0)\right]
-
\nn
-
\left[B(x_0)+C(x_0)\right]U(x_0,y_0)\hat G_H(x_0,y_0)
=0.
\ee

When postulating that $U(x_0,y_0)$ satisfies the differential equation

\be
\frac{d}{dx_0}U(x_0,y_0)-B(x)U(x_0,y_0)=0,
\label{link}
\ee

the above expression reduces to:

\be
U(x_0,y_0)\left[\frac{d}{dx_0}\hat G_H(x_0,y_0)\right]
-
C(x_0)U(x_0,y_0)\hat G_H(x_0,y_0)
=0.
\ee

This formula can be transformed by multiplying with $U^{-1}(x_0,y_0)$ from the
left:

\be
\left[\frac{d}{dx_0}\hat G_H(x_0,y_0)\right]
-
U^{-1}(x_0,y_0)C(x_0)U(x_0,y_0)\hat G_H(x_0,y_0)
=0.
\ee

As has been shown above, the solutions for the last equation and equation 
(\ref{link}) are given by

\be
\hat G(x_0,y_0)
=
\p\left\{\int^{x_0}_{y_0}d\xi_0
U^{-1}(\xi_0,y_0)C(\xi_0)U(\xi_0,y_0)
\right\}
\ee

and

\be
U(x_0,y_0)
=
\p\left\{\int^{x_0}_{y_0}d\xi_0
B(\xi_0) 
\right\},
\ee

respectively. Making use of the group property (\ref{addth}) the last line 
of equation (\ref{general}) is reproduced. These derivations can be repeated
for any variable $n\cdot x$. Only for $n^2=0$ the Dirac equation cannot be written in
the generic form (\ref{genericdirac}).


\end{document}